\documentclass[journal,10pt]{IEEEtran}
\usepackage{xcolor}
\usepackage{relsize}

\usepackage{ifpdf}
\usepackage{graphicx}
\usepackage{cite}
\usepackage{nomencl}
\usepackage{enumitem}
\usepackage{slashbox}
\usepackage{amsmath}
\usepackage[normalem]{ulem}
\usepackage{algorithm}
\usepackage{algorithmic}

\ifCLASSINFOpdf
\else
\fi
\usepackage{amsmath,amssymb,bm}
\usepackage{algorithmic} 
\usepackage{adjustbox}
\usepackage{array}
\usepackage[tablename=Table]{caption}
\usepackage{stfloats}
\usepackage{url}
\usepackage{amsthm}
\usepackage{multirow}

\newcommand{\RomanNumeralCaps}[1]{\MakeUppercase{\romannumeral #1}}
\usepackage{stackengine,scalerel}

\newcommand\overstar[1]{\ThisStyle{\ensurestackMath{%
			\setbox0=\hbox{$\SavedStyle#1$}%
			\stackengine{0pt}{\copy0}{\kern.2\ht0\smash{\SavedStyle*}}{O}{c}{F}{T}{S}}}}

\newcommand{\quotes}[1]{``#1''}

\DeclareMathOperator*{\argmin}{argmin} 
\begin{document}
 
\title{Bayesian Graph Neural Network for Fast identification of critical nodes in Uncertain Complex Networks}

\author{\IEEEauthorblockN{Sai Munikoti,~\textit{Student Member, IEEE}, Laya Das, Balasubramaniam Natarajan,~\textit{Senior Member, IEEE}
		\thanks{This work has been submitted to the IEEE for possible publication. Copyright may be transferred without notice, after which this version may no longer be accessible. Preprint submitted to IEEE SMC 2021}
		}}


\maketitle

\begin{abstract}
In the quest to improve efficiency, interdependence and complexity are becoming defining characteristics of modern complex networks representing engineered and natural systems. 
Graph theory is a widely used framework for modeling such complex networks and to evaluate their robustness to disruptions. Particularly, identification of critical nodes/links in a graph can facilitate the enhancement of graph (system) robustness and characterize crucial factors of system performance. Most existing methods of critical node identification are based on an iterative approach that explores each node/link of a graph. These methods suffer from high computational complexity and the resulting analysis is network specific. Additionally, uncertainty associated with the underlying graphical model further limits the potential value of these traditional approaches. To overcome these challenges, we propose a Bayesian graph neural network based node classification framework that is computationally efficient and systematically incorporates uncertainties. Instead of utilizing the observed graph for training the model, a MAP estimate of the graph is computed based on the observed topology and node target labels. Further, a Monte-Carlo (MC) dropout algorithm is incorporated to account for the epistemic uncertainty. The fidelity and the gain in computational complexity offered by the Bayesian framework is illustrated using simulation results.
\end{abstract}

\IEEEpeerreviewmaketitle

\section{Introduction}


Graph theory is a popular framework to study complex systems and is gaining popularity with the growth of network size, complexity and data availability. Hence, various graph specific machine learning algorithms are recently being proposed in the literature for applications like node classification, community detection, link prediction, etc. \cite{lu2020lstm,bongini2021molecular}. The proper functioning of a complex system depends on its constituent sub-systems (nodes) and their interconnections (links). Usually, there exists a set of critical nodes that are relatively more crucial to the overall operation of system. The removal of critical nodes severely disrupts the network functioning. The network functionality is often studied in the literature with the help of graph robustness metrics such as effective graph resistance, flow robustness, etc. \cite{alenazi2015comprehensive}. Robustness quantifies the impact of loss of resources (nodes) on the performance of a system. Due to the inherent topological structure, each node contributes differently to graph robustness, i.e., its removal/loss affects the robustness to a different degree. In this regard, the notion of node criticality score is introduced, which quantifies the decrease in robustness (such as effective graph resistance) when the corresponding node is removed from the graph \cite{wang2014improving, munikoti2020inductive}.
Critical scores are then employed to classify the nodes, with the high criticality assigned to the nodes whose removal maximally decreases the graph robustness.  

There are numerous methods proposed in the literature for identifying critical nodes \cite{boginski2009identifying, wang2014improving, wang2015network,van2011decreasing}. However, there are four major drawbacks with these conventional approaches of identifying critical nodes: (1) Computationally expensive due to iterative nature of the algorithms; (2) The results obtained from the iterative approaches are specific to the network considered and cannot be generalized to other systems/graphs or alternate robustness metrics; (3) Topological information is not explicitly used in node identification; (4) Accuracy of the approximation methods decreases as the size of the network grows. To overcome these challenges, we have proposed a graph neural network based scalable framework in \cite{munikoti2020inductive} that efficiently computes criticality scores and consequently identifies critical nodes/links in large complex networks. Although \cite{munikoti2020inductive} can address some of the fundamental challenges of existing approaches, it still overlooks one of the fundamental problems of graph-based models, i.e., the uncertainty in the underlying graph (system). Most of the current works on node criticality consider the observed graph as a perfect model of the underlying complex system. However, in reality, this is not always the case. For instance, a node (user) in a social network influences another node with some probability. In protein-protein interaction networks, the edges, representing interactions among proteins, are results of noisy and error-prone measurements. 

Uncertainties are present in the graph because of modeling assumptions, insufficient data and noisy measurements. Uncertainties can lead to the addition of spurious links/weights in the graph or miss the important links between the nodes that have a strong relationship. Consequently, any robustness analysis of the observed graph could lead to inaccurate inferences. There are a few prior attempts to incorporate the uncertainties while developing graphical models of the systems. For instance, authors in \cite{lam2018modeling} have used fuzzy set theory to determine the link weights. However, the approach in \cite{lam2018modeling} is not very generic. From a learning based perspective, authors in \cite{zhang2019bayesian, pal2019bayesian} have extended the classic Graph Neural Network (GNN) to a Bayesian framework. Here, the observed graph is viewed as a realization of a parametric random graph model, and then the joint posterior estimate of graph parameters as well as GNN weights is computed. 

Inspired by this work, we propose a novel Bayesian inductive learner for graph robustness (BILGR). The nodes are classified into three classes based on the criticality scores. BILGR is trained on subset of nodes and then prediction can be done for any node of the graph. The proposed node classification approach is scalable, computationally efficient, and systematically accounts for the uncertainties in link connections.
To the best of the authors' knowledge, this is the first attempt to develop a Bayesian framework of critical node identification through GNN.

\vspace{-0.5cm}
\subsection{Related work}
The authors in \cite{wang2014improving} explore effective graph resistance as a measure of robustness for complex networks. Robustness is improved by protecting the node/link whose removal maximally increases the effective graph resistance. The complexity of an exhaustive search to identify such a link is of order $O(N^5)$ for a graph of $N$ nodes. To overcome computational complexity, the authors in \cite{pizzuti2018genetic}, propose a method based on genetic algorithm to enhance network robustness. Particularly, the authors focus on identifying links whose removal would severely decrease the effective graph resistance of the graph. However, the algorithm requires manual fine-tuning of various parameters and is not scalable. Authors in \cite{munikoti2020inductive} proposed a learning based framework for fast identification of critical nodes/links in large graphs. However, the framework does not handle graph uncertainties. Specifically for GNN, authors in \cite{zhang2019bayesian, pal2019bayesian}, leverage a Bayesian framework to systematically incorporate uncertainties. The fundamental idea behind the proposed method is similar to the Bayesian neural network where some prior distribution is assumed for the unknown weights of the network and then the posterior is computed for a different set of weight parameters sampled from Markov chain Monte Carlo (MCMC) algorithm. From various experiments, the Bayesian GNN formulation shows that, although computationally more demanding, it leads to (1) an ability to learn more rapidly with less data, (2) a better capacity to represent uncertainty, and (3) better robustness and resilience to noise or adversarial attacks.
\subsection{Contributions}
This work develops a new framework for estimating node criticality class while systematically accounting for uncertainties. The key contributions of this work are as follows:
\begin{itemize}
    \item  A novel critical node identification framework is developed with a GNN-based inductive learning approach that is applicable for graphs in the presence of aleatoric and epistemic uncertainties.
    \item The scalability of the proposed technique is demonstrated by training GNN models on a small subset of nodes and predicting node classes in a relatively larger fraction of the graphs that were not used for training.
    \item Systematic integration of aleatoric and epistemic uncertainty through MAP estimate of underlying graph and Monte-Carlo (MC) dropout algorithm, respectively.
    \item The superior performance is validated on power law cluster graphs with more than $90$\% identification accuracy and credible intervals. The computational efficiency is demonstrated with execution time three orders smaller compared to the conventional approaches. 
\end{itemize}

The rest of the paper is organized as follows. Section \RomanNumeralCaps{2} describes the fundamentals of classic GNN followed by Bayesian GNN and graph robustness metrics. Section \RomanNumeralCaps{3} presents the proposed framework with experiments and results in Section \RomanNumeralCaps{4}. Final conclusions are provided in Section \RomanNumeralCaps{5}.

\section{Background}
This section summarizes the working principle of classic GNN and Bayesian GNN, which are then employed to devise our BILGR framework.  
\subsection{Fundamentals of GNN}
Recently, new generalizations and operations have been developed for extending deep learning approaches to graph based data.
\cite{kipf2016semi} is one of the first works that efficiently extended the two dimensional convolutional operations to graph convolutional operations. In GNNs, the foremost task is to represent a node by a vector such that it captures both structural as well as the desired target information. These node embedding vectors can then be used for various downstream applications like node classification, link prediction, by appending the node embeddings with the classic feed forward layers. The standard procedure to learn this embedding vector from initial node features involves a \quotes{ message passing mechanism}. Here, the information (node feature) is aggregated from the neighbors of a node and then combined with its own feature to generate a new feature vector. This process is repeated to generate the final embedding for each node of a graph.

Broadly, there are two methodologies to learn node embeddings. The first category belongs to transductive based learning algorithms like deepwalk of \cite{hamilton2017inductive} and node2vec, where a complete graph is needed to learn the embeddings of nodes. 
On the contrary, the second category belongs to an inductive based learning algorithm, namely the GraphSAGE \cite{hamilton2017inductive}. This algorithm learns the mapping (aggregator) function instead of learning the embedding vectors. 
Hence, GraphSAGE can induce the embedding of a new node or node unseen during training, given its features and neighborhood. 
GraphSAGE learns a representation for every node based on some combination of its neighboring nodes, parametrized by $K$. 
The parameter $K$ controls the number of hops to be considered as a neighbor. 
After defining the neighborhood, the aggregate function is employed to collect neighbor's embedding for a selected node. The aggregate function could be as simple as a mean function or as complex as a multilayer perceptron with trainable weights. 
For initialization, the node feature vectors are assigned as embeddings. Thereafter, for each neighborhood depth until $K$, a neighborhood embedding is generated with the aggregator function for each node and concatenated with the existing node embedding. Passing the concatenated vector through a feed forward multilayer layer perceptron (MLP) layer results in the final node embeddings. GraphSAGE learns the weights of an aggregator function and MLP by minimizing a relevant loss function. 
In a supervised learning setup, the algorithm tries to reduce the loss associated with node target labels.
Once weights are learned, then an embedding vector for any node can be predicted given the node features and its neighboring information. Our framework is based on GraphSAGE as it is more appropriate for large dynamic graphs.

\subsection{Fundamentals of Bayesian GNN}
Bayesian GNN is an extension of GNN to account for uncertainties in the underlying graph.
It is possible to take a Bayesian perspective on GNN by constructing a joint posterior distribution of the graph, the
GNN weights and the node labels. The marginal posterior probability of node labels is given by,
\begin{equation}
\begin{split}
p(\bm{Z|Y_{L}}, \bm{X},G_{obs}) = \int p(\bm{Z|W,X},G_{obs}) ,\\
 p(\bm{W|Y_{L},X},G_{obs}) p(G|G_{obs}, \bm{X, Y_{L}})d\bm{W}dG,
\end{split}
\label{eq:2}
\end{equation}
where $\bm{W}$ denotes the random weights of a Bayesian GNN over graph $G$ \cite{pal2019bayesian}. As the integral in (\ref{eq:2}) cannot be computed analytically, MCMC approximation is formed as,
\begin{equation}
\begin{split}
p(\bm{Z|Y_{L}}, \bm{X},G_{obs}) \approx \frac{1}{S}\sum_{s=1}^{S}\frac{1}{N_{G}}\sum_{i=1}^{N_{G}}
\int p(\bm{Z|W_{s,i},X}, G_{obs})
\end{split}
\label{eq:3}
\end{equation}
In this approximation, $N_{G}$ graphs $G_{i}$ are sampled from $p(G|G_{obs},\bm{X,Y_{L}})$ and $S$ weight samples
$W_{s,i}$ are drawn from $p(\bm{W|Y_{L},X,G_{i}})$ by training a Bayesian GNN corresponding to the graph $G_{i}$. The final prediction result is an average of all the predictions coming from different samples. Here, one can sample a number of similar distributions but diverse topologies based on the posterior of the graph generation model. Thus, in the aggregation step, the GNN model is able to learn more general node embeddings by incorporating information from different potential neighbors.

\subsection{Robustness metrics}
Robustness of a graph signifies the resistance of a graph against node/link failure. This is typically measured via the connectivity of the graph post failure. 
Thus, higher robustness implies higher connectivity of the residual graph post node/link failure. 
There are various network-based surrogate metrics that approximately quantify the graph robustness against node failure \cite{alenazi2015comprehensive, munikoti2021robustness}. The authors in \cite{alenazi2015comprehensive, alenazi2015evaluation } studied and compared various graph theoretic metrics for approximating network robustness. 
The key take away from \cite{alenazi2015comprehensive, alenazi2015evaluation } is that no single metric can approximate the robustness for all the different families of graph under all scenarios. Thus, different metrics are used based on the study and type of graphs. Effective graph resistance ($R_{g}$) appears to be most accurate and generic metric for graph robustness \cite{wang2014improving}. Therefore, we focus on $R_{g}$ in this study. It is important to note that the proposed approach can be extended to other surrogate robustness metrics. $R_{g}$ is the sum of the effective resistances over all pairs of vertices \cite{ellens2011effective}. 
$R_{g}$ considers both the number of paths between the nodes and their length (link weight), intuitively measuring the presence and quality of back-up possibilities. $R_{g}$ can be computed in several ways and equation (\ref{eq:1}) shows the spectral form.
\begin{equation}
    R_{g} = \frac{2}{N-1}\sum_{i=1}^{N-c}\frac{1}{\lambda_{i}},
    \label{eq:1}
\end{equation}
where $\lambda_{i}$ are the eigen values of the laplacian matrix of graph $G$ and $c$ is the number of connected components in the graph, which matches with the number of zero eigen values. Thus, only non-zero eigen values are utilized while computing $R_{g}$. 

\section{Proposed approach}
The proposed BILGR has two key features: (1) It incorporates data (i.e., graph) and model (i.e., GNN) uncertainty. (2) It poses identification of critical nodes as node classification problem where classes denote the criticality level of the nodes. The framework is described in the following steps.

\subsection*{\textit{Step 1- Train node embeddings}}
The embedding vector for nodes is generated by maximally utilizing all the available information namely the observed graph $G_{obs}$, initial feature vector $\bm{X}$, and targets $\bm{Y_{L}}$. Here, targets denote the node criticality score/class. Algorithm 1 summarizes the steps involved in computing node criticality scores \cite{munikoti2020inductive}. The criticality scores are used to classify nodes into three classes as shown in the Table \ref{Table:class labels}. High criticality nodes correspond to class 1. Each node is assigned a feature vector $\bm{X}$, consisting of node weighted degree as well as its average neighbor degree. 

In contrast to \cite{zhang2019bayesian}, here we propose to learn the embeddings in a supervised way by training a node classification model on $G_{obs}$ with node class labels.
The GNN based classification model consists of a GraphSAGE layer with depth $K=3$. Once the GraphSAGE layer aggregates features from the neighbors, a feed forward layer is connected to the output followed by a softmax activation. The detail architecture is shown in figure \ref{fig:2}.
Thereafter, the intermediate output of the trained model, i.e., the final output of graph aggregation operation ( before the feedforward layer) would serve as node embeddings.
\begin{table}[h!]
    \centering
    \caption{Class labels of nodes based on criticality scores}
    \label{Table:class labels}
	\begin{tabular}{|c|c|}
	\hline
    Criticality Score & class label \\ 
    \hline
    $0-0.3$ & Class 1 \\
    \hline
    $0.3-0.7$ & Class 2 \\
    \hline
    $0.7-1.0$ & Class 3 \\
    \hline
  \end{tabular}
\end{table}

\subsection*{\textit{Step 2-Compute node distance matrix}}
The node distance matrix represents the connectivity of nodes in an embedding space. In other words, it determines the relative position of nodes by computing the distances between them. The distance between any two arbitrary nodes is \cite{zhang2019bayesian}, 
\begin{equation}
\begin{split}
Z_{i,j}(\bm{X}, G_{obs}) = ||\pmb{e_{i}-e_{j}}||^2,
\end{split}
\label{eq:4}
\end{equation} 
where, the vectors $\pmb{e_{i}}$ and $\pmb{e_{j}}$ are the embedding vector of node $i$ and $j$, respectively.

\subsection*{\textit{Step 3-Obtain MAP estimate of graph}}
In real world modeling scenarios, the underlying graph may not be known accurately. There are uncertainties associated with the link connections or link weights. This type of uncertainty in an underlying graph belongs to the category of an aleatoric uncertainty. Assuming the observed graph $G_{obs}$ is a sample from a collection of random graph models, we attempt to a get a posterior distribution of graph, i.e., $P(G|G_{obs}, \bm{X, Y_{L}})$. This posterior distribution is then plugged into equation ($1$) to generate final node predictions. The analytical derivation of probability distribution is intractable. To counter the intractability, one can design a suitable Markov chain Monte Carlo (MCMC) simulation to sample from the posterior. However, MCMC is challenging and computationally expensive for the case of graph, especially when we are dealing with thousands and millions of nodes. Therefore, a maximum a posteriori estimate (MAP) of the graph is obtained. Here, the term estimate of a graph refers to estimate of the weighted adjacency matrix of the graph. If the embedding vectors of the graph nodes are assumed to be a signal residing on a graph, then one can use the framework of learning a graph from the smooth signals. The smoothness criterion will of the estimated graph. The graph  
$G_{est}$ (equivalently $A_{G_{est}}$) is computed by solving the following graph learning optimization algorithm \cite{ zhang2019bayesian, kalofolias2017large},
\begin{equation} 
A_{G_{est}} = \argmin_{A_{g}} ||A_{G} \circ Z||- \alpha\bm{1}^T\log(A_{G}\bm{1}) + \beta||A_{G}||_{F}^{2} \\
\label{eq:5}
\end{equation} 
where, $A_{G}$ is the set of feasible symmetric matrices; $Z$ is the distance matrix as computed in Step 2. The Hadamard product of $A_G$ and $Z$, i.e., $A_G \circ Z$, represents the Dirichlet energy of a graph, which quantifies the smoothness of the signals on a weighted graph. 
The second term of the optimization problem in equation (\ref{eq:5}) ensures that the node degree is never zero although the individual link weights can be zero. The last term of the equation constrains the adjacency matrix to be sparse. In a nutshell, the optimization problem tries to learn a sparse weighted adjacency matrix that is a better representation of given node embedding vectors. \\

\subsubsection*{\textit{Step 4-Train GNN over the inferred graph $G_{est}$ and employ MCMC to sample different GNN weights}}
Once a graph $G_{est}$ is estimated, a new node classification model is trained on $G_{est}$ with node class labels. This trained model is the final model that would predict class labels for test nodes. 
However, here the GNN model yields point predictions and does not account for epistemic uncertainty. Therefore, a Monte Carlo dropout algorithm is implemented to account for epistemic uncertainty \cite{gal2016dropout}. In training time, we run multiple forward passes with different dropout masks each time. During prediction, we collect different predictions of test samples with different dropout masks. Then, mean and variance of these predictions are used to devise the predictive posterior mean and credible interval, respectively. 

\begin{figure*}
    \includegraphics[height=6.5cm, width=18cm]{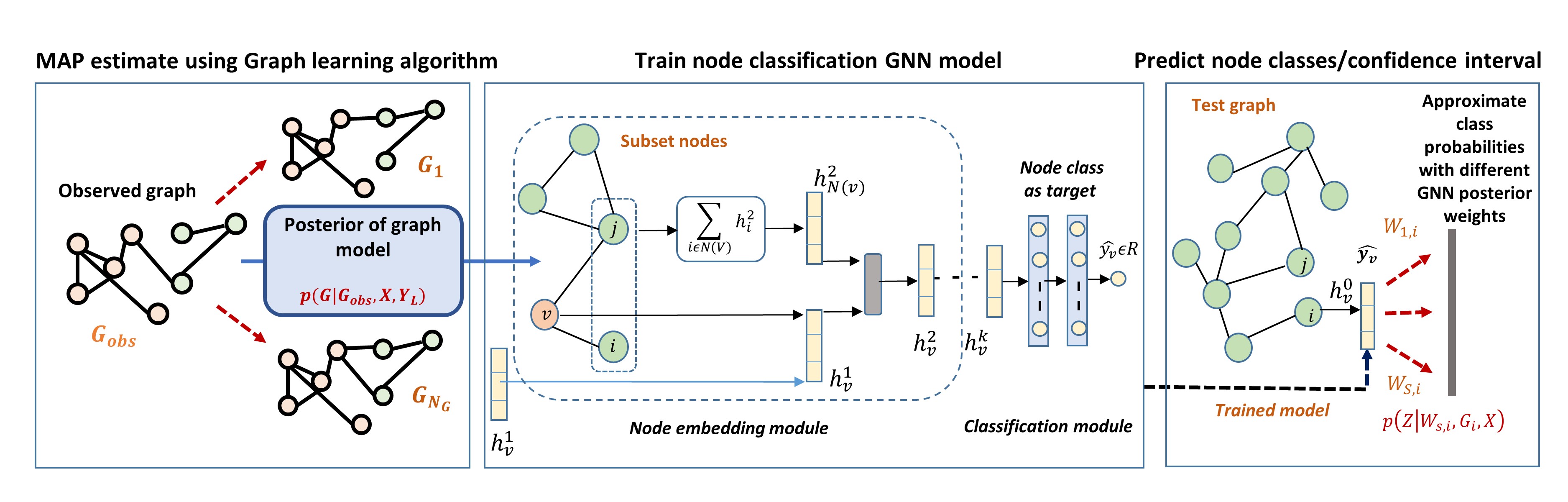}
    \caption{Proposed BILGR framework for critical node identification under graph/model uncertainty}
    \label{fig:2}
\end{figure*}

Algorithm 2 summarizes all the steps from 1-4.
The schematic in figure \ref{fig:2} describes the complete Bayesian framework of GNN for critical node identification. The details of model training and performance evaluation experiments is discussed in the next section.

\section{Experiments}

\subsection{Baseline approach}
To evaluate the performance of our proposed approach, we compare the BILGR with a conventional method of identifying node criticality. A typical procedure involves removing a particular node from the graph and computing the robustness metric of the residual graph. This process is repeated for all the nodes of the graph. Thereafter, the metrics of all nodes are used to generate node ranks and identify the high criticality nodes whose removal maximally decreases the graph robustness. Algorithm $1$ summarizes the existing approach of identifying high criticality nodes.

\begin{algorithm}[h!]
 \caption{Conventional algorithm of critical node identification}
 \begin{algorithmic}[1]
 \renewcommand{\algorithmicrequire}{\textbf{Input:}}
 \renewcommand{\algorithmicensure}{\textbf{Output:}}
 \REQUIRE Graph $G$ with $V$ nodes.
 \ENSURE  Node class based on criticality score 
 \\ \textit{LOOP Process}
  \FOR { $n$  in \textit{V}}
  \STATE  Remove node $n$ from graph \textit{G} 
  \STATE  Compute robustness metric of the residual graph ($G-n$)
  \STATE Assign criticality scores to node $n$
  \ENDFOR
 \RETURN Assign class to nodes based on criticality scores (refer Table \ref{Table:class labels})
 \end{algorithmic} 
 \end{algorithm}
 
 \begin{algorithm}[h!]
 \caption{Propose BILGR algorithm to identify critical nodes}
 \begin{algorithmic}[1]
 \renewcommand{\algorithmicrequire}{\textbf{Input:}}
 \renewcommand{\algorithmicensure}{\textbf{Output:}}
 \REQUIRE Observed graph $G_{obs}$ with node feature vector $\bm{X}$ and subset node class $\bm{Y_{L}}$.
 \ENSURE  Node class, i.e. node criticality.  
 \STATE Learn node embeddings in a supervised manner using $G_{obs}$, $\bm{X}$  and $\bm{Y_{L}}$.
 \STATE Compute node distance matrix using eqn. (\ref{eq:4}).
\STATE Solve optimization problem in eqn. (\ref{eq:5}) to obtain adjacency matrix of MAP estimate of graph
\STATE Sample multiple networks weights using MC dropout masks by training model on the estimated graph.
\STATE Predict multiple class probabilities for each test nodes using different training dropout masks.  
 \RETURN mean and variance of predictions to get posterior mean and credible interval of node classes. 
 \end{algorithmic} 
 \end{algorithm}
 
\subsection{Model settings and Training:} 

There are various hyper-parameters in the model that need to be tuned for obtaining the best possible model. The number of node embedding layers, i.e., the depth of the GNN is selected as three. The number of neurons in these three layers are $64$, $32$, and $16$ respectively. The classification module consists of two feedforward layers with $10$ and $3$ neurons respectively. The activation function in all the layers is kept as \textit{relu} except last layer which has softmax. The loss function is categorical cross entropy which is optimized via ADAM optimizer. The training of GNN and Graph generation/manipulation are carried out in TensorFlow and python NetworkX, respectively.  
  
\subsection{Results}
The node identification problem is framed as node classification problem where three classes denote the three levels of node criticality, i.e., high, medium and low. However, in any graph, critical nodes is only $2\%$ to $5$ $\%$ of the total nodes, thereby inducing a class imbalance problem. The imbalance is addressed by employing weighted version of GNN also known as cost sensitive learning. In the loss function, the weights are assigned to a class proportional to the inverse of the class distribution present in the training dataset. The weights are $100$, $50$ and $1$ for classes $1$,$2$ and $3$, respectively. The performance of the models is reported with accuracy, precision and recall. Here, recall is an important metric as we want to correctly predict all class-1 nodes, i.e., high criticality nodes which are usually small in number. 

The model is trained on subset of $3000$ nodes in a power law cluster graph of $5000$ nodes and $5997$ edges. Power law cluster is a family of graphs that follow a power law degree distribution and also exhibit \quotes{clustering} by requiring that, in some fraction of cases ($p$), a new node connects to a random selection of the neighbors of the node to which the new node was connected last time. This graph family is selected for illustration as many real-world networks have shown to topologically resemble it. However, our framework is generic for any family of the graph as well as any robustness metric. Table \ref{Table:Obs} tabulates the scores of the trained model when evaluated on a unseen $2000$ test nodes. As we are more interested in the class $1$ performance since they represent high criticality nodes, the scores are tabulated for class-1 as well. It can be seen that the trained model is $90$ \% accurate in detecting all the classes of nodes. More importantly, the model is correctly identifying $95$ \% of the top critical nodes (i.e., $41$ out of $43$) with high recall score of $0.95$.
\begin{table}
    \centering
    \caption{Performance of the GNN model}
    \label{Table:Obs}
	\begin{tabular}{|c|c|c|}
	\hline
    Model & Accuracy  & Recall  \\ 
    \hline
    Overall & 0.902 $\pm$ 0.009 & 0.900    \\
    \hline
    Class-1 & 0.953  $\pm$ 0.01 & 0.950  \\
    \hline
  \end{tabular}
\end{table}

Apart from high identification accuracy, the proposed BILGR framework offers significant advantage in execution time compared to a conventional iterative approach. Specifically, BILGR identifies $43$ high criticality nodes in $17$ seconds compared to $64600$ seconds with a conventional approach in a $5000$ nodes graph. Thus, the proposed  method is multiple orders faster than the conventional  approach,  and  this  gap  will increase  as  the network  size  grows. The  training  and  experiments  are conducted on a system with an Intel i7 processor running at 3.4GHz with 6GB Nvidia RTX 2070 GPU.

\begin{figure}
    \includegraphics[height=5cm, width=9cm]{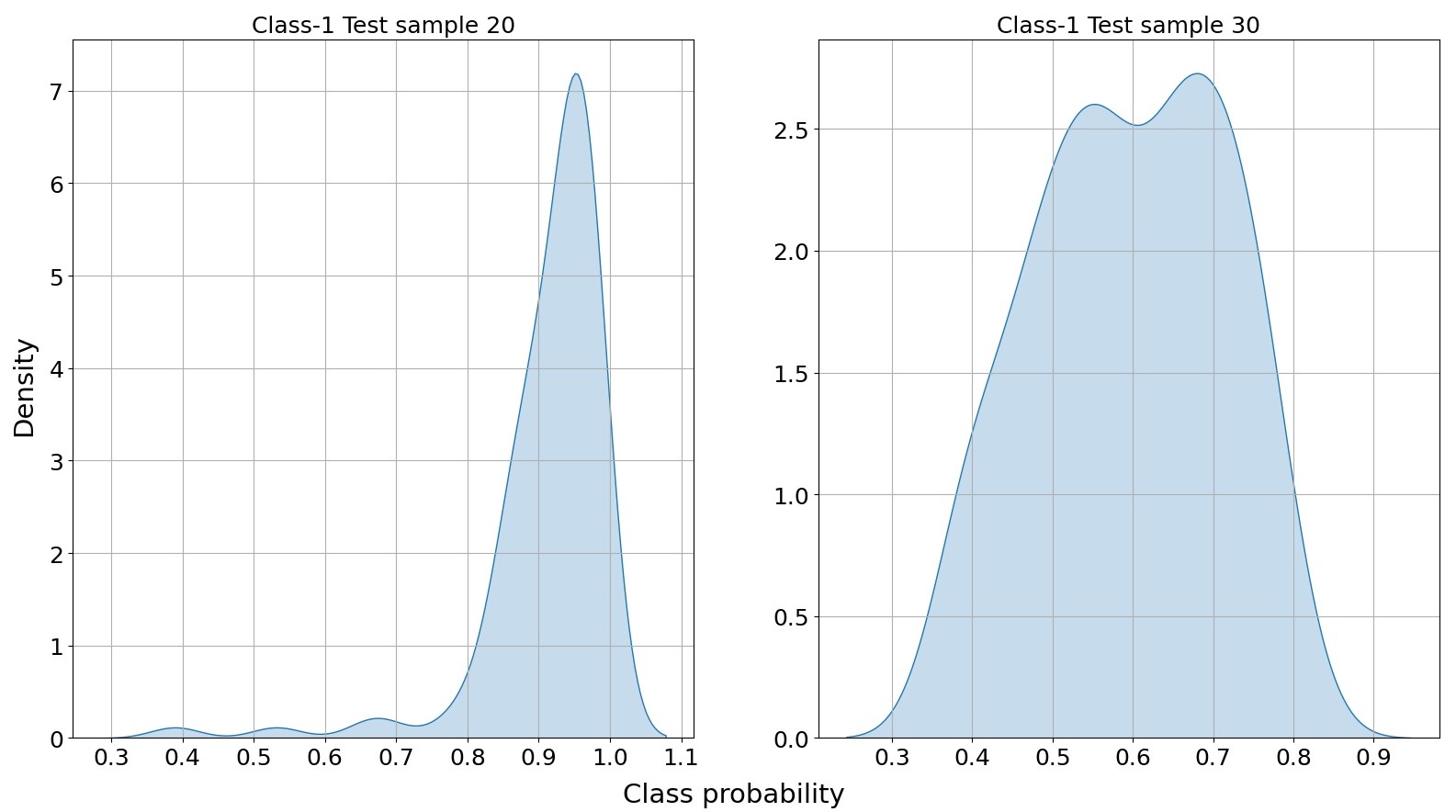}
    \caption{Probability distributions of class probabilities for test samples of class-1 }
    \label{fig:4}
\end{figure}

The Bayesian formulation of the proposed framework provides confidence in the predictions which is useful when test nodes differs significantly from the training nodes in terms of topological properties such as neighborhood sub-graph, assortativity, etc. The confidence around the mean predictions of accuracy is shown in the Table \ref{Table:Obs}. Further, to demonstrate the applicability of confidence interval, the prediction of two  class-1 test samples are shown in the   
Figure \ref{fig:4}. It can be observed that the model is more confident with test sample $1$ (left hand side) with a tighter confidence interval compared to test sample $2$. This type of extra inference is only possible in Bayesian framework, which further increases the usability of the model. 

Due to the systematic incorporation of graph and model uncertainty, the model is also robust to noisy or adversarial graphs. To demonstrate this, noisy versions of graph are generated by randomly adding links at $5$\% of the nodes. Table \ref{Table:Gest_Robustness} lists the scores of the model when evaluated on two noisy graphs (i.e., Gnoise1 and Gnoise2) with $6096$ and $6196$ links. The number of new randomly added links in Gnoise1 and Gnoise2 are $100$ and $200$, respectively. It can be observed that model prediction remains consistent to a very high degree even though spurious links are added to the input graph. This demonstrates the robustness of the BILGR framework to noisy or adversarial test samples (nodes).   
Furthermore, with MC dropout algorithm we took $100$ predictions of each test node and reported the posterior mean and confidence interval of the predictions in tables \ref{Table:Obs} and \ref{Table:Gest_Robustness}. Compared to point prediction, the accuracy with ensemble prediction has increased by approx $1$\% for each of the test scenarios.

\begin{table}[h!]
\centering
\caption{Robustness of Trained model to graph noise}
\begin{tabular}{|c|c|c|c|}
\hline
\textbf{Test graph}                                                                    & \textbf{Measure}  & \textbf{Accuracy} & \textbf{Recall} \\ \hline
\multirow{2}{*}{\textbf{\begin{tabular}[c]{@{}c@{}}Gnoise1\\ 6096 links\end{tabular}}} & \textbf{Ovverall} & 0.896 $\pm$ 0.09        & 0.866           \\ \cline{2-4} 
                                                                                       & \textbf{Class-1}  & 0.931 $\pm$ 0.01        & 0.930           \\ \hline
\multirow{2}{*}{\textbf{\begin{tabular}[c]{@{}c@{}}Gnoise2\\ 6196 links\end{tabular}}} & \textbf{Ovverall} & 0.876$\pm$0.08        & 0.886           \\ \cline{2-4} 
                                                                                       & \textbf{Class-1}           & 0.890$\pm$0.02        & 0.89            \\ \hline
\end{tabular}
\label{Table:Gest_Robustness}
\end{table}

As mentioned earlier, all the results are demonstrated for power law cluster graph. A different model is needed for each family of synthetic graphs, i.e., power law, power law cluster, Stochastic block model, among others. This is because, different families of graphs vary in their overall structure and link connections, i.e., degree distributions, assortativity, average clustering coefficient, etc. However, model trained on one family can be employed for prediction of other graphs (even larger one) of the same family. Indeed, models trained on synthetic graphs can be used for identification of critical nodes in real networks \cite{munikoti2020inductive}. This is possible due to the generalizability of our BILGR framework ( inductive learning) and topological resemblance of real networks with synthetic graphs.

\section{Conclusions}
This work proposes a Graph neural network based BILGR framework to quickly identify critical nodes under uncertain conditions. The criticality is related to node robustness metrics such as effective graph resistance, weighted robustness, among others. BILGR addresses data uncertainty by computing a MAP estimate of a graph. Then, a GNN architecture based on node classification is designed with node labels belonging to one of the three classes. The model is trained on the estimated graph and tested on unseen nodes. To incorporate model uncertainty, MC dropout is included in the GNN architecture. The model has sufficient accuracy to identify critical nodes as well as it provides a confidence interval around its predictions. The significant computational advantage over classical approach of critical node identification is demonstrated through algorithm running times. As a part of future work, the BILGR will be extended to critical Link identification and other graph families. 

\section*{Acknowledgement}
This material is based upon work supported by National Science Foundation under award number $1855216$.
\bibliographystyle{IEEEtran}
\bibliography{IEEEabrv, BILGR}

\end{document}